\documentclass{aa}
\usepackage[varg]{txfonts}
\usepackage{graphicx}
\usepackage{xcolor}
\usepackage{multirow}
\usepackage{multicol}
\usepackage{txfonts}
\usepackage[normalem]{ulem}

\def\gsim{\,\lower4pt\hbox{${\buildrel\displaystyle >\over\sim}$}\,}
\def\lsim{\,\lower4pt\hbox{${\buildrel\displaystyle <\over\sim}$}\,}

\usepackage{textcomp,gensymb} 

\begin{document}

\title{Particle energization in colliding subcritical collisionless shocks investigated in the laboratory}
   
\author{A. Fazzini\inst{1}, W. Yao\inst{1,2}, K. Burdonov\inst{1,2,3}, J. B\'eard\inst{4}, S. N. Chen\inst{5}, A. Ciardi\inst{2}, E. d'Humi\`eres\inst{6}, R. Diab\inst{1}, E.~D.~Filippov\inst{3,7}, S. Kisyov\inst{5,\thanks{Present address: Lawrence Livermore National Laboratory, Livermore, California 94550, USA}}, V. Lelasseux\inst{1}, M. Miceli\inst{8,9}, Q. Moreno\inst{6,10}, S. Orlando\inst{9}, S. Pikuz\inst{7,11}, X. Ribeyre\inst{6}, M. Starodubtsev\inst{3}, R. Zemskov\inst{3}, J. Fuchs\inst{1}}

\institute{LULI - CNRS, CEA, UPMC Univ Paris 06 : Sorbonne Universit\'e, Ecole Polytechnique, Institut Polytechnique de Paris - F-91128 Palaiseau cedex, France 
\and
Sorbonne Universit\'e, Observatoire de Paris, Universit\'e PSL, CNRS, LERMA, F-75005, Paris, France
\and
IAP, Russian Academy of Sciences, 603950, Nizhny Novgorod, Russia
\and
LNCMI-T, CNRS, 31400, Toulouse, France
\and
“Horia Hulubei” National Institute for Physics and Nuclear Engineering, RO-077125 Bucharest-Magurele, Romania
\and
University of Bordeaux, Centre Lasers Intenses et Applications, CNRS, CEA, UMR 5107, F-33405 Talence, France
\and
JIHT, Russian Academy of Sciences, 125412, Moscow, Russia
\and
Università degli Studi di Palermo, Dipartimento di Fisica e Chimica E. Segrè, 90134, Palermo, Italy
\and
INAF, Osservatorio Astronomico di Palermo, 90134, Palermo, Italy
\and
ELI-Beamlines, Institute of Physics, Czech Academy of Sciences, 5 Kvetna 835, 25241 Dolni Brezany, Czech Republic
\and
NRNU MEPhI, 115409, Moscow, Russia
}

\date{\today}
 
  \abstract
   {
   Colliding collisionless shocks appear in a great variety of astrophysical phenomena and are thought to be possible sources of particle acceleration in the Universe.}
   {The main goal of our experimental and computational work is to understand what is the effect of the interpenetration between two subcritical collisionless shocks
   on particle energization.
   }
   {To investigate the detailed dynamics of this phenomenon, we have performed a dedicated laboratory experiment.
   We have generated two counter-streaming subcritical collisionless magnetized shocks by irradiating two teflon (CF$_2$) targets with 100 J, 1 ns laser beams on the LULI2000 laser facility.
   The interaction region between the plasma flows was pre-filled with a low density background hydrogen plasma and initialized with an externally applied homogeneous magnetic field perpendicular to the shocks.
   We have also modeled the macroscopic evolution of the system via hydrodynamic simulations and the microphysics at play during the interaction via Particle-In-Cell simulations.}
   {We report here on measurements of the plasma density and temperature during the formation of the supercritical shocks, their transition to subcritical, and final interpenetration.
   We found that in the presence of two shocks the ambient ions reach energies around 1.5 times of the ones obtained with single shocks.
   Both the presence of the downstream zone of the second shock and of the downstream zone common for the two shocks play a role in the different energization: the characteristics of the perpendicular electric fields in the two areas allow, indeed, certain particles to keep being accelerated or to avoid being decelerated.
   }
   {The findings of our laboratory investigation are relevant for our understanding of the energy distribution of high-energy particles that populate the interplanetary space in our solar system and the very local interstellar medium around the heliopause, where the observations have found evidence of subcritical collisionless shocks that may, eventually, collide with each other.}


\titlerunning{Counter-streaming collisionless subcritical shocks}
\authorrunning{A. Fazzini et~al.}

\maketitle
\section{Introduction}

Subcritical collisionless shocks are a class of shocks that are able to satisfy the Rankine-Hugoniot jump conditions using only dispersive and resistive dissipation mechanisms (\citealt{balogh2013physics}).
In subcritical shocks, the downstream flow velocity exceeds the sound speed behind the shock, but is  lower than the magnetosonic speed (defined as $c_{ms} = \sqrt{c_s^2 + v_A^2}$, where, $c_s$ and $v_A$ are the ion sound velocity and Alfv\'{e}nic velocity, respectively). The limiting pre-shock magnetosonic Mach number $M_{ms} = v_{s} / c_{ms}$ for these conditions to be satisfied depends on the $\beta$ of the plasma and on the shock obliquity (i.e., the angle between the shock velocity and the upstream magnetic field). Subcritical shocks thus need to satisfy $M_{ms} \lesssim M^{cr}_{ms}$, with $M^{cr}_{ms}$ ranging between $M^{cr}_{ms} = 1$  for quasi-parallel shocks, up to $M^{cr}_{ms} = 2.76$ for perpendicular shocks, in the limit $\beta\rightarrow0$ (\citealt{Edmiston1984}).
In astrophysics one can find these so-called subcritical shocks in a variety of scenarios. When a high-Mach number flow meets a dense medium, it becomes heavily ``mass-loaded'' and slows down to a velocity that allows the formation of a subcritical shock. This is expected to happen when the solar wind interacts with the interstellar medium and forms the termination shock (\citealt{Treumann2009}).  Moreover, some astrophysical supercritical shocks evolve into subcritical ones in the course of their interaction with the upstream medium and consequent loss of energy, as it happens for solar coronal mass ejections (CME; \citealt{Bemporad2011}). The collision of subcritical shocks are expected to occur between forward and reverse shocks in the solar wind and also between solar wind shocks and planetary bow shocks (\citealt{whang1985}).\\
Like their counterpart, namely supercritical shocks with $M_{ms} > M_{ms}^{cr}$, where ions can be accelerated through a variety of mechanism (\citealt{balogh2013physics,Marcowith2016}), subcritical shocks can also accelerate ions and induce thermal heating, although particle acceleration does not play a significant dissipative role in subcritical shocks. Both ion acceleration and heating have been observed in satellite crossings (\citealt{Mellot1984}).  The underlying ion acceleration mechanism(s) is still up for debate, but it is suggested to include $\mathbf{v} \times \mathbf{B}$ heating (\citealt{Ohsawa1985}), ion reflection to a small degree from the shock front (\citealt{lee1987}), and from other wave-particle interactions (\citealt{Balikhin1996}). As for particle acceleration from the collision of two subcritical shocks, no significant ion acceleration, with respect to the energies reached by particles accelerated by supercritical shocks, was observed in simulations (\citealt{cargill1986interaction}).
Accelerated ions with energy in the tens of MeV have been measured from the collision of two subcritical shocks at a small angle (\citealt{dudkin2000simulation}), however their numbers are extremely small and there is still an ongoing effort to determine the acceleration mechanism.\\
Recently, high-power lasers and externally controlled magnetic field generation have opened the door to investigations of astrophysically relevant collisionless shock studies on particle acceleration (\citealt{Li2019,Fiuza2020,yao2021laboratory,yao2021detailed}). \\
Here, we created in the laboratory subcritical perpendicular collisionless shocks, i.e., inside an external magnetic field perpendicular to the shock propagation direction. 
We characterize in detail their global spatio-temporal dynamics using multiple diagnostics. Moreover, we investigate the head-on encounter of two such shocks, in order to determine if and how this could modify the conditions under which ions can be energized in such a configuration. The shocks were characterized in the laboratory by interferometry and Thomson scattering (TS) measurements, performed at different times, which provided the electron density map, local electron density, and local ion and electron temperature. We then performed three-dimensional (3D) hydrodynamic simulations with the magneto-hydrodynamic (MHD) code FLASH (\citealt{FLASH}), which reproduced the global dynamics of both the expanding plasmas driving the shock, as well as the latter.
Next, we studied the event using the 1D3V fully kinetic Particle-In-Cell code SMILEI (\citealt{derouillat2018smilei}) where we used again the experimentally obtained parameters as the initialization values. In these kinetic simulations, we observed particle acceleration of the ambient particles.
The acceleration is initially due  to the electrostatic field associated with the shock front $E_x$, then to the inductive electric field $E_y \sim v_x B_z$, where $v_x$ is the flow velocity and $B_z$ the perpendicular magnetic field. During the interaction between the two subcritical shocks, we note that both the presence of the downstream zone of the second shock and the creation of a downstream zone common for the two shocks play a role in the higher energization of the ambient ions: the characteristics of the perpendicular electric fields of these two areas allow, indeed, certain protons to keep being accelerated or to avoid being decelerated. As a result, ambient ions were energized to 1.5 times the energy of the single shock case. This is consistent with space measurements performed in-situ of ions accelerated ahead of outward propagating interplanetary shocks (\citealt{Gosling1984}), or in the interaction of an interplanetary shock with the bow shock of the Earth (\citealt{hietala2011situ}).

\section{Experiment}

\subsection{Setup and diagnostics}
\begin{figure}[htp]
    \centering
    \resizebox{\hsize}{!}{\includegraphics[width=1.\linewidth]{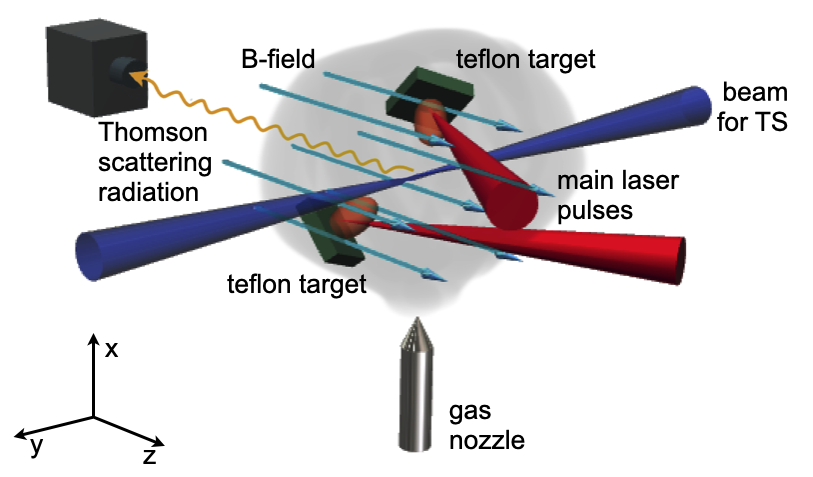}}
    \caption{Setup of the experiment, conducted at LULI2000 by having two high-power lasers (1 ns, 100 J at 1$\omega$, $1.6\times 10^{13}$ W/cm$^2$ on target) irradiate two solid (Teflon, CF$_2$) targets to investigate the interpenetration of two magnetized shocks. An auxiliary beam of ~15 J was used to perform Thomson scattering (TS) and an additional low energy beam (not shown in the picture for readability reasons) probed the plasma along a line titled 9º upwards with respect to the z-axis in order to measure the integrated plasma electron density.}
    \label{fig:double_shock_setup}
\end{figure}

\begin{figure}[htp]
    \centering
    \resizebox{\hsize}{!}{\includegraphics[width=0.7\linewidth]{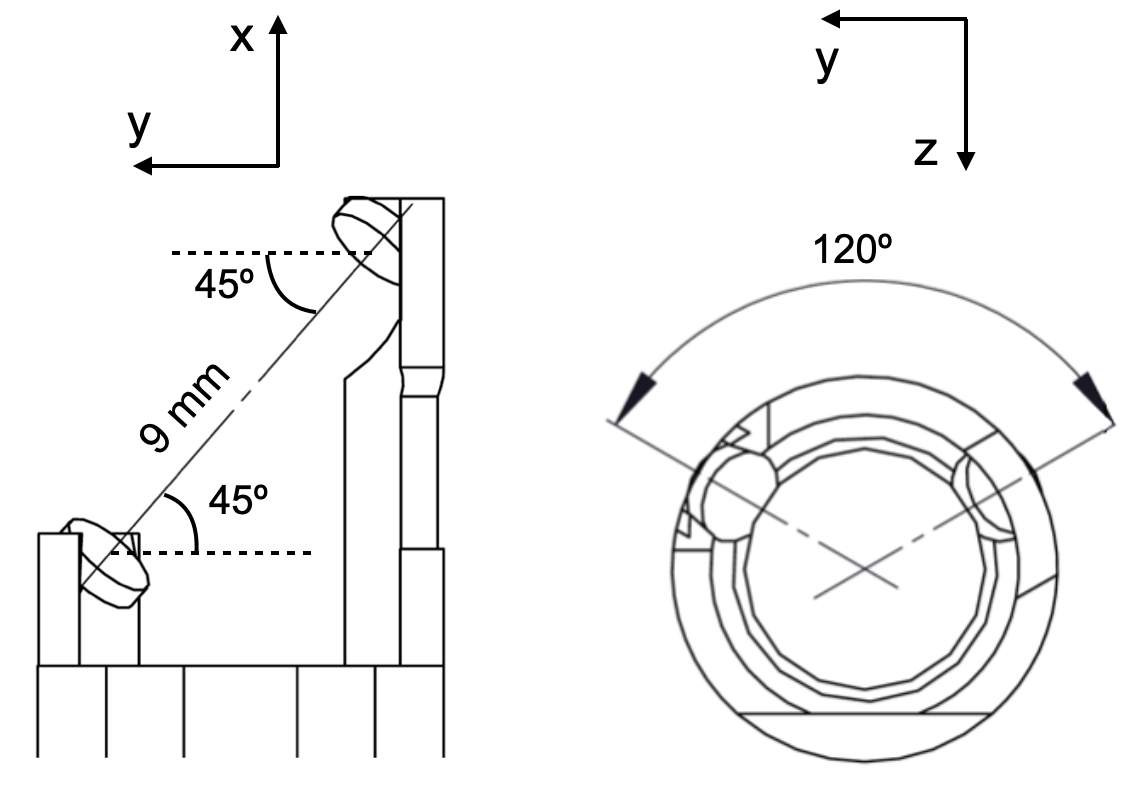}}
    \caption{Side (left) and top (right) view of the targets. }
    \label{fig:targets}
\end{figure}

The setup employed in our experiment is shown in Fig.~\ref{fig:double_shock_setup}: we irradiated two teflon (CF$_2$) targets with two high-power laser pulses  (1053 nm wavelength, 1 ns, 100 J, $1.6 \times 10^{13}$ W/cm$^2$ each). The targets were tilted in a way that allowed the laser beams to reach them and such that the two plasma flows would encounter each other, as detailed in Fig.~\ref{fig:targets}. The two targets were separated by a 9 mm distance. The region in between the targets was pre-filled with hydrogen at low density ($n_0 \sim 10^{18}$ cm$^{-3}$) injected by a gas nozzle and magnetized using an externally applied magnetic field of 20 T provided by a pulsed coil (\citealt{albertazzi2013production}), directed along the z-axis.\\
A focusing spectrometer with spatial resolution (FSSR) (\citealt{Faenov1994}) was utilized to register x-ray ion emission of the plasma with and without ambient medium. It allowed to characterize both the plasma initiated by the laser interaction at the surface of each target, and the heating of the ambient medium induced by the expanding plasmas. The spectrometer was equipped with a spherically bent mica crystal with a lattice spacing  $2d = 19.9149$~\r{A} and curvature radius of $R=150$~mm. It was able to measure He-like (transitions 3p--1s, 4p--1s, 5p--1s etc.) and H-like (transition 2p--1s and its satellites) lines of Fluorine in the range of wavelengths between 13 and 16~\r{A} with a spatial resolution about 0.1~mm along the axis which joins the centers of both targets (see the blue line in Fig.~\ref{fig:interf_2shocks}-d).
The presence of Sulfur impurities in the targets allowed us also to register a corresponding He-like doublet (2p--1s transition) in the third order of reflection with a Li-like satellite structure being sensitive in our range of plasma parameters. Spectral resolution was achieved better than $\lambda/d\lambda=1000$. The spectra were recorded using Fujifilm Image Plates of type TR, which were placed in a cassette holder protected from the visible optical radiation. The signal is time-integrated. The analysis of x-ray spectra was done by comparison of the experimental line ratios with simulated ones using the radiative-collisional code PrismSPECT (\citealt{Macfarlane2004}) and by comparison of emissivity profiles in different conditions.
\\
A high-energy auxiliary beam (527.5 nm wavelength, 1 ns, 15 J, focused over $\sim$ 40 \textmu m along the y-axis and propagated throughout the plasma, see Fig.~\ref{fig:double_shock_setup}) was used to perform Thomson scattering (TS) measurements off the electron and ion waves in the plasma. It was used in a mode where the plasma was sampled in a collective mode (\citealt{froula2011plasma}). The collection of the scattered light was performed at 90$\degree$ (along the z-axis) from the incident direction of the laser probe (the y-axis).
The light scattered off the ion (TSi) and electron (TSe) waves in the plasma was analyzed by means of two different spectrometers, set to different dispersions (3.1 mm/nm for TSi and $7.5 \times 10^{-2}$ mm/nm for TSe), which were coupled to two streak-cameras (Hamamatsu for TSe, and TitanLabs for TSi, both equipped with S-20 photocathode to be sensitive in the visible part of the spectrum, and both with typical 30 ps temporal resolution), allowing us to analyze the evolution of the TS emission in time.
The scattering volumes sampled by the instruments were: 
120 \textmu m along the x- and y-axes, 40 \textmu m along the z-axis for TSi; 100 \textmu m along the x- and y-axes, 40 \textmu m along the z-axis for TSe.
The analysis of the Thomson scattered light was performed by comparison of the experimental images (recorded by the streak cameras) with the theoretical curves of the scattered spectrum for coherent TS in non-collisional plasmas, with the instrumental function width of 5.9 nm for the electron spectrometer and 0.12 nm for the ion spectrometer taken into account.
We point out that the TS laser probe induces some heating in the hydrogen ambient gas (details can be found in \citealt{yao2021laboratory}).
With TS, we can get a spatially and temporally resolved measurement of the plasma density and temperature. 
In addition, another optical probe beam ($\lambda = 530$ nm) passed with a 9$\degree$ angle with respect to the B-field lines through the interaction zone, allowing for a measurement of the integrated electron density through interferometry.

\subsection{Experimental results}

\begin{figure}[htp]
    \centering
    \resizebox{\hsize}{!}{\includegraphics[width=0.48\textwidth]{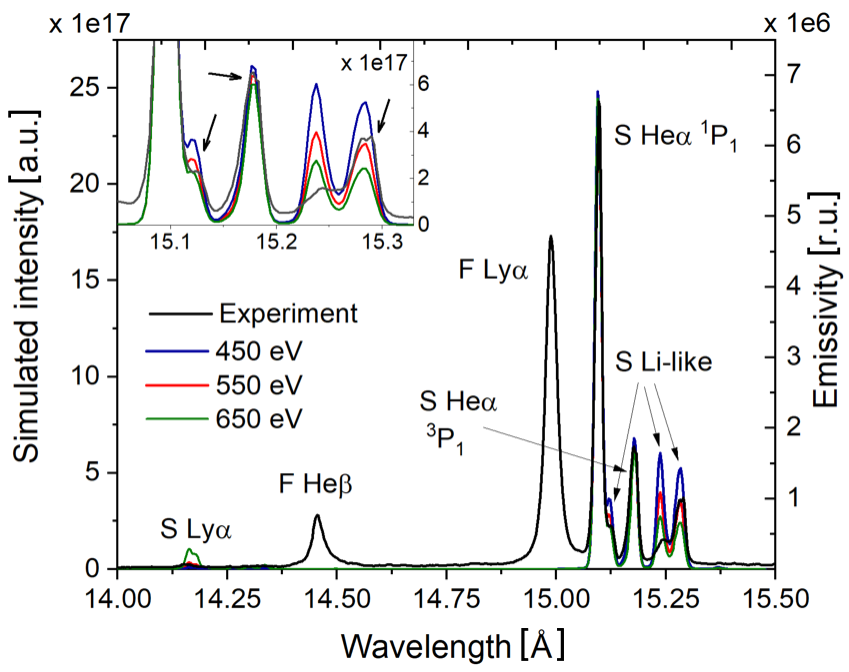}}
    \caption{Experimental x-ray spectrum (black line) measured by the FSSR spectrometer as emitted from a CF$_2$ target. What is recorded is the spectrum of Sulfur impurities in the third order of reflection. Overlaid are simulations performed using the PrismSPECT code (red, olive and blue curves) for the target surface region, using the group of satellites sensitive to the plasma parameters. For all temperatures shown in the figure, the electron density was $N_e=7 \times 10^{20}$~cm$^{-3}$. All curves are normalized to the S He$_{\rm \alpha}$ line. The best fitting corresponds to the red curve. The inset  shown in the top left corner  demonstrates the detailed fitting of the satellites of the experimental spectrum. The arrows point to the lines having the best fit.  
    }
    \label{fig:FSSR-surface}
\end{figure}

The electron temperature on the target surface was measured via FSSR by recording the emission of Sulfur lines and by simulating this emission  in a steady-state approach using the code PrismSPECT (\citealt{Macfarlane2004}). This is shown in Fig.~\ref{fig:FSSR-surface}, yielding for the surface plasma a temperature $T_e=550$~eV at almost critical density $N_e=7 \times 10^{20}$~cm$^{-3}$. We point out that this measurement is relative to the laser-target interaction, i.e., to the collisional part of the system.\\
\begin{figure}[htp]
    \centering
    \includegraphics[width=0.42\textwidth]{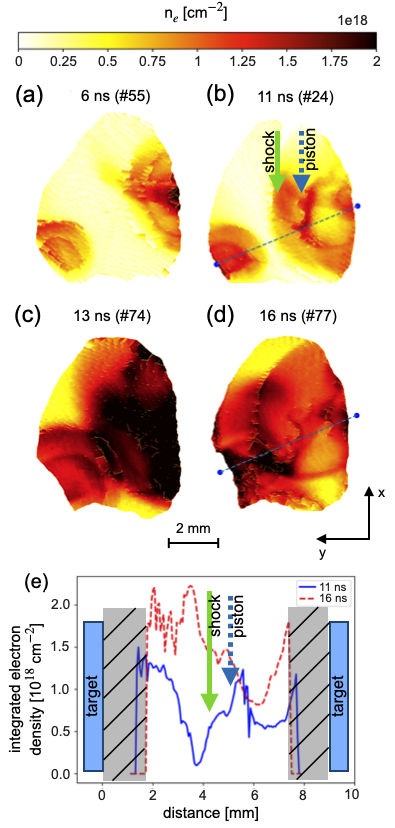} 
    \caption{Temporal sequence of integrated (along the z-axis, with a 9$\degree$ tilt) electron density measurements showing the evolution of the interpenetration of two magnetized shocks, at 6 ns (a), 11 ns (b), 13 ns (c), and 16 ns (d), after the main laser pulses hit the targets. The aperture of the magnetic coil structure restrained the passage of the optical probe, diminishing slightly the field of view (FoV) in (a-d). (e) shows the lineout of the integrated electron density at 11 ns (blue solid line) and at 16 ns (red dashed line), along the lines shown on the relative map - (b) and (d), respectively; the location of the targets at 9 mm distance from each other is also shown, while the gray dashed areas represents the zones out of the FoV. Before the interpenetration of the two plasma flows, a shock front and a piston develop out of each target, as indicated by the two arrows in (b) and (e): the dotted blue arrow points at the left-drifting piston and the solid green arrow at the left-drifting shock front. When the two shocks collide, the density has a spike at the meeting point at around 3.5 mm in (e).
}
    \label{fig:interf_2shocks}
\end{figure}
After the plasmas have been generated at the surface of each target, they expand into the ambient medium. This expansion is monitored by optical probing. This is displayed in Fig.~\ref{fig:interf_2shocks}, which shows the measurements, at successive times, of the integrated (along the line-of-sight of the probe beam) electron density of the plasmas expanding from both targets. We point out that these images were obtained on different shots.
As seen in our previous experiment where we created one single magnetized shock (\citealt{yao2021laboratory}),  two structures develop out of each target: a piston front and a shock front characterized by two separated bumps of higher electron density (identified by arrows in Fig.~\ref{fig:interf_2shocks}b and Fig.~\ref{fig:interf_2shocks}e).
Each piston is the result of the expansion of the plasma ablated from the solid target by each laser.
The plasma flows expand in the low-density ambient hydrogen, which is quickly ionized by the x-rays produced by the irradiated targets, and shocks are generated as a result of the combined action of the supersonic piston expansion and of the externally applied B-field.
In fact, the strong external magnetic field of 20 T is critical in providing additional pressure so that a magnetized shock can form in the hydrogen plasma (\citealt{yao2021laboratory}), as in its absence we would get a shock only in the presence of a denser background plasma. 
As a result, for early times, i.e., before $\sim$ 12 ns, we observe two well-developed shocks propagating against each other. \\
In our situation, the shocks are perpendicular, i.e., the angle between the magnetic field and the shock propagation direction is $\theta_{Bn} \approx 90 \degree$, and are characterized by a $\beta = P_{therm}/ P_{mag} \approx 0.1$, hence the critical Mach number has a value $M_{ms}^{cr} \sim 2.6$ (\citealt{Edmiston1984}). 
The shocks obtained in our experiment are supercritical up to 3-4 ns after the laser beams hit the targets and turn into subcritical for later times (\citealt{yao2021laboratory}). Indeed, they propagate with an initial velocity of $v_s \approx$ 1500 km/s, which corresponds to $M_{ms} \approx$ 3.3 > $M_{ms}^{cr}$, and when they eventually interact, they have a velocity of a few hundreds of km/s, which gives, for $v_s \approx 500$ km/s, $M_{ms} \approx 1.1 < M_{ms}^{cr}$.
We point out that the measurements of shock velocity are obtained from the interferograms by measuring the positions at different times, which correspond also to different shots.
As we can observe, the structures developing from the two targets have different sizes: they start forming at the same time, but their different distance from the gas nozzle exhaust makes them propagate in a medium of slightly different density which has a visible impact on their propagation velocities.
Moreover, the fact that the two shocks do not propagate directly against each other, but perpendicularly to the targets, has also been taken into account while calculating the velocity. Indeed, the interferometry view corresponds to the side view (apart for a $\sim9\degree$) of Fig.~\ref{fig:targets}, which is a projection of the displacements along the z-axis. Hence, the distances extracted from the interferometry figures have been multiplied by a factor of $1 / \cos (60 \degree) = 2$.

As for the collisionality, we find that the mean-free-path of the drifting ions with respect to the ambient ones is $\lambda_{mfp}^{i-i\, (d-a)} \approx 33$ mm (calculated according to \cite{Braginskii1965}), which is much larger than the characteristic length over which the interaction takes place ($\sim$ hundreds of \textmu m) and hence makes the shock collisionless.
\begin{figure}[htp]
    \centering
    \resizebox{\hsize}{!}{\includegraphics[width=0.48\textwidth]{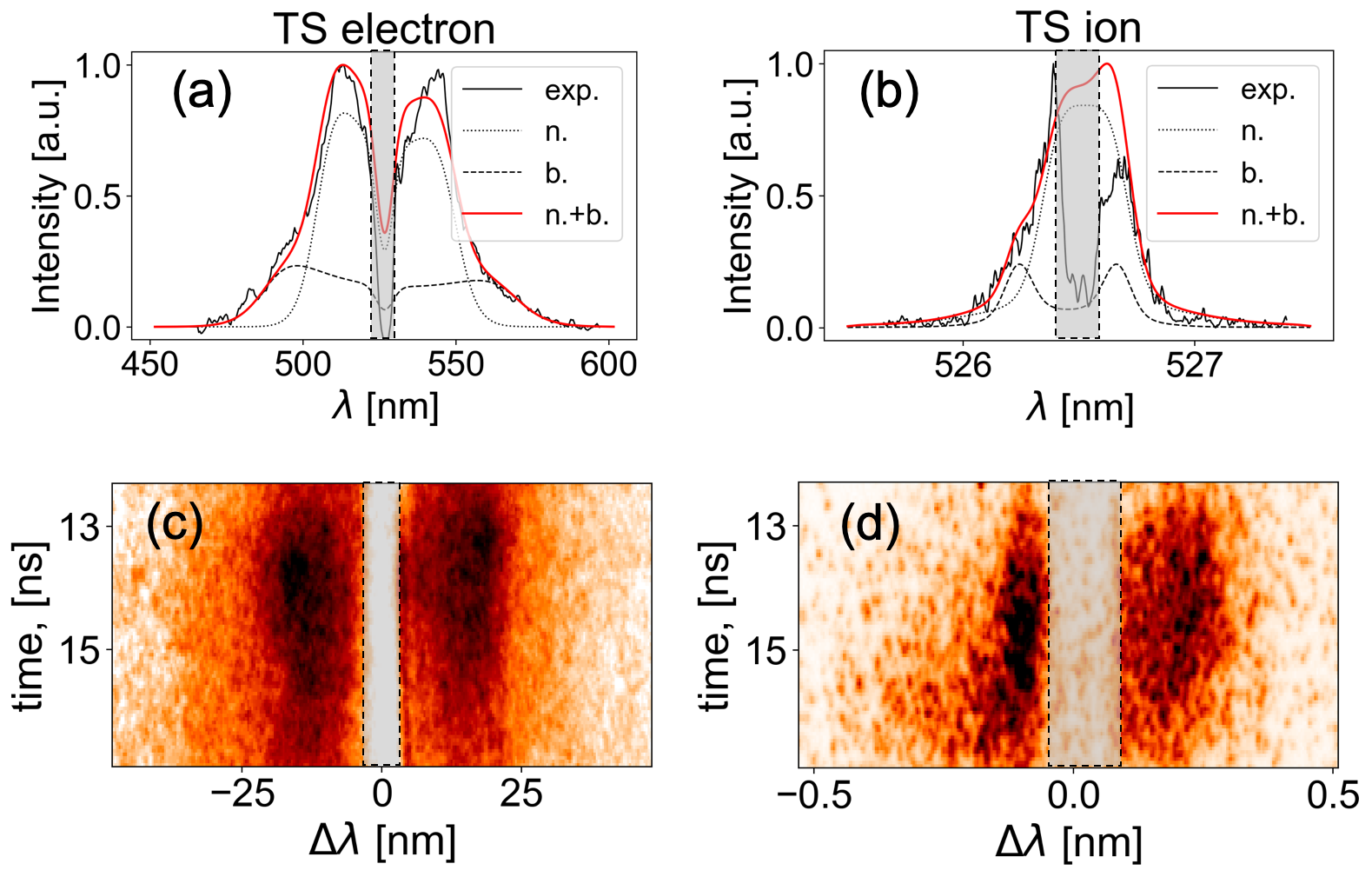}}
    \caption{Thomson scattering measurements of the plasma density and temperatures in the region of shock collision. Spectra of Thomson scattering off electron plasma waves ((a),(c)) and ion acoustic waves ((b),(d)). (a) and (b) show the spectra profiles, corresponding to 15 ns, while (c) and (d) show the temporal evolution of the scattering spectra over a time period from 13 ns to 16 ns for the electron plasma and ion acoustic waves, correspondingly. 
    Black solid lines (in (a) and (b)) are for experimental data profiles, while red solid lines are for theoretical spectra, composed of a superposition of narrow (black dotted lines) component relative to the ambient medium, having  density 1.5 $\times 10^{18}$ cm$^3$, electron temperature 100 eV, ion temperature 200 eV, and broad (black dashed lines) component relative to the piston plasma, having density 6 $\times 10^{18}$ cm$^3$, electron temperature 300 eV, ion temperature 100 eV. The ratio between the magnitudes of the narrow and broad components is 3.5. We note that the deep central dip in the experimental spectra ((a),(b)) and the white vertical region in the streak-camera images ((c),(d)) is related to a filter (a black aluminum stripe) which is positioned right before the entrance of the two streak cameras (recording respectively the light scattered off the electron and ion waves). This filter is used to block the very intense and unshifted laser wavelength (the Rayleigh-scattered light), which otherwise would saturate the cameras. Thus, no signal is recorded in this zone, which is materialized by the grey dashed box.
    }
    \label{fig:ts_double_shock_result}
\end{figure}

\begin{figure}[htp]
    \centering
    \includegraphics[width=0.44\textwidth]{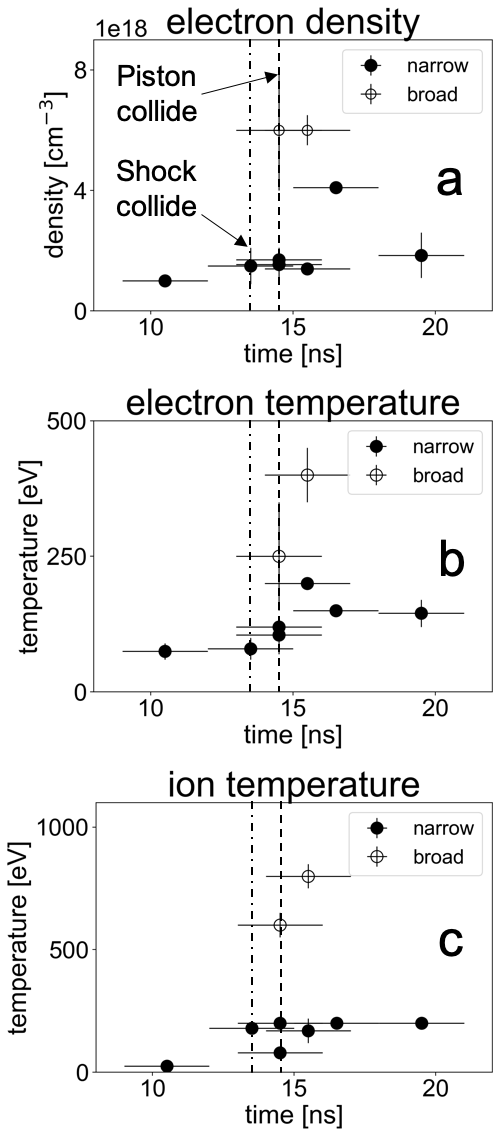}
    \caption{TS measurements of the temporal evolution of electron density (a), electron temperature (b), and ion temperature (c). The narrow and broad configurations are respectively related to the ambient and piston plasmas (see caption of Fig.~\ref{fig:ts_double_shock_result}).}
    \label{fig:ts_graphs}
\end{figure}

Moreover, we measured the plasma Thomson scattering of the plasma thermal waves to assess the plasma characteristics.
Fig.~\ref{fig:ts_double_shock_result} shows two examples of TS spectra from electron plasma waves ((a) and (c)) and from ion acoustic waves ((b) and (d)), corresponding to 15 ns and the period from 13 ns to 16 ns, respectively.
The temporal evolution of the electron density, the electron temperature, and ion temperature is shown in Fig.~\ref{fig:ts_graphs}.
We observe that after around 13 ns the TSi signal suddenly broadens, which corresponds to the time of the collision between the two shocks. This broadening is attributed to the heating of the ions in the plasma due to the energy released when the two plasma bubbles collide. As is shown in the time evolution in Fig.~\ref{fig:ts_graphs}, after the collision, the electron density slightly increases, while the electron temperature remains initially unperturbed (around 80 eV). The interpenetration of the two plasma shocks heats the ions up to temperatures $\approx 135$ eV, according to an adiabatic gas compression.
Electrons are then heated at a slower rate by ion-electron collisions. 
Further increase of ion and electron temperature as well as electron density is observed when the pistons collide at a later time ($\sim 14.5$ ns). Here, we will only focus on the shock-shock collision, before the encounter of the pistons.\\
\begin{figure}[htp]
    \centering
    \resizebox{\hsize}{!}{\includegraphics[width=0.48\textwidth]{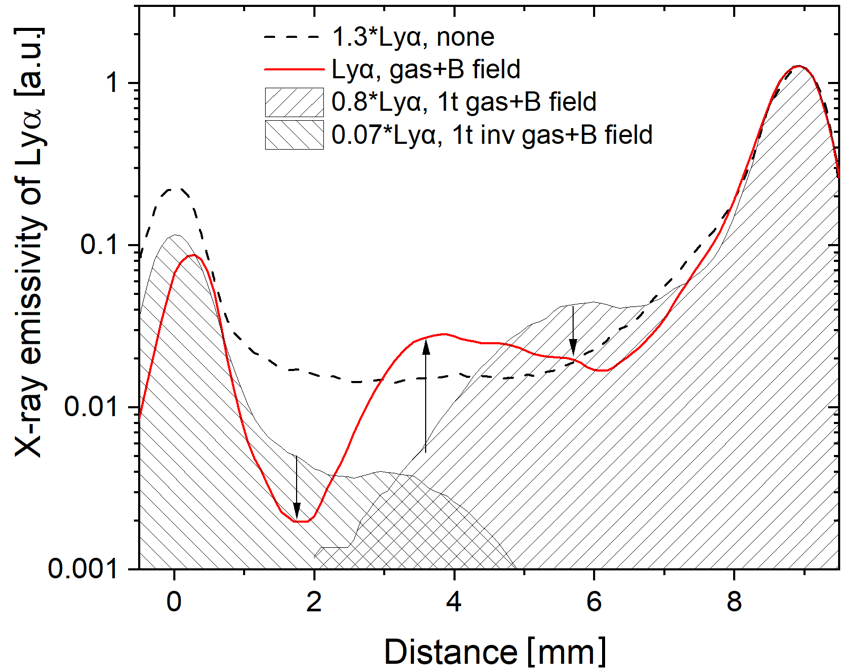}}
    \caption{X-ray emissivity profiles measured by the FSSR spectrometer in different cases. First when two plasmas expand from the two targets, either in the absence (dashed black line) or presence (red line) of the ambient medium. 
    The gray curves with a pattern correspond to a single target case as a reference. In all cases, the magnetic field is present. All curves are normalized to the right target emissivity. The emissivity for the left target is the inverted right one with a multiplier taking into account the signal reduction due to the location of a part of the spectrometer in the shadow zone.
    }
    \label{fig:FSSR-emissivity}
\end{figure}

\begin{table}
\begin{center}
\caption{List of parameters extracted from our measurements at $\sim 11$ ns, i.e., right before the interpenetration of the two shock structures. $\lambda^{i-i \, (d-a)}_{mfp}$ is the collisional mean free path between drifting and ambient ions.}
\label{tab:params}
\begin{tabular}{cc}
\hline
\multicolumn{2}{c}{\textbf{Characterized Ambient Plasma Conditions}}    \\
\hline
Upstream Elec. Number Density $n_e$ {[}cm$^{-3}${]}                      & $1.0 \times 10^{18}$\\
Upstream Elec. Temperature $T_e$ {[}eV{]}     & 80     \\
Upstream Ion Temperature $T_i$ {[}eV{]}      & 20     \\
Downstream Elec. Temperature $T_e$ {[}eV{]}  & 130    \\
Downstream Ion Temperature $T_i$ {[}eV{]}    & 200      \\
Shock Velocity at meeting point $v_s$ [km/s]   & $\sim 500$       \\
Upstream Magnetic Field Strength $B_z$ {[}T{]}  &  20      \\ 
\hline
\multicolumn{2}{c}{\textbf{Calculated Parameters}}      \\
\hline
Ion Collisional mean-free-path $\lambda_{mfp}^{i-i \, (d-a)}$ [mm]    & 33   \\
Flow Ion Larmor Radius $r_{L,i,fl}$	[mm]	&		0.26	 \\
Upstream plasma Thermal Beta $\beta_{ther}$          & 0.10   \\
Mach Number $M$       & 4.42                  \\
Alfv\'enic Mach Number $M_A$ &       1.15            \\
Magnetosonic Mach Number $M_{ms}$         &     1.12             \\
 \hline
\end{tabular}
\end{center}
\end{table}

Complementary to the TS measurements, Fig.~\ref{fig:FSSR-emissivity} shows the x-ray emissivity profiles of the plasma located in between the two targets.
We compare three  cases: (1) when the applied  magnetic field and the ambient medium are present and the plasmas expand from the two targets (thin red line), (2) still with two plasmas, but in the absence of the ambient medium  (dashed black line), and (3) when only one plasma is flowing from either the right or the left target, but in the same magnetic field and ambient medium conditions as in case (1) (thin gray lines, with areas  filled by  patterns).
What we observe is that the collision between the two plasma flows results in an emission enhancement in the zone between the two targets (compare the red curve to the filled areas). Here the left target has a lower intensity due to the positioning of the corresponding part of the spectrometer in the "shadow zone" of the right target. The electron temperature was measured in this region as $T_e=240$~eV (a lower-limit estimate) at an electron density $n_e=10^{18}$~cm$^{-3}$ using the ratio between the resonance lines Ly$_{\rm \alpha}$ and He$_{\rm \beta}$ by the method described in (\citealt{Khiar2019}), in reasonable agreement with the TS measurements, knowing that the x-ray diagnostics is time-integrated. In addition, one can note that the emissivity drops significantly faster between the target and the middle zone in the case when the ambient medium is applied which is most probably related to a faster recombination rate as well as to a higher confinement of particles close to the target (\citealt{Filippov2021}). \\

The main plasma parameters extracted from the experimental measurements are summarized in Table~\ref{tab:params}.
These values are used to initialize our simulations detailed below, which we use to further investigate the particle acceleration during the shock collision.

\section{Numerical simulations}

Our simulation effort is two-fold: the first step was to undertake MHD simulations of the laser-driven plasma expansion and interaction with the ambient gas and magnetic field, leading to the experimentally observed piston. In the second step, we have used the results of the experimental diagnostics as a starting point for kinetic simulations that allow us to investigate in details the microphysics of the shocks colliding and the underlying particle acceleration mechanisms.

\subsection{MHD simulations}
\begin{figure}[htp]
    \centering
    \resizebox{\hsize}{!}{\includegraphics[width=0.48\textwidth]{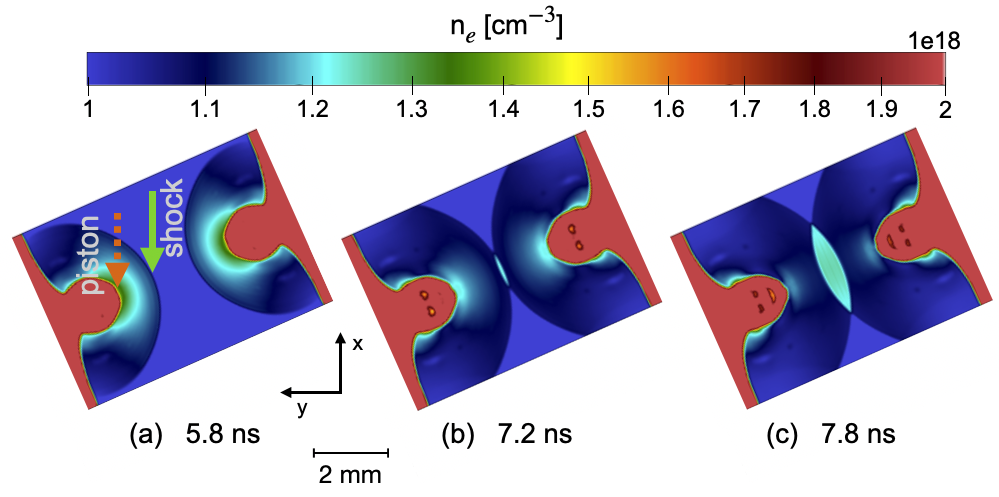}}
    \caption{Simulation, using the 3D MHD code FLASH, of the volumetric electron density plotted at $t=5.8$ ns (a), $t=7.2$ ns (b), and $t=7.8$ ns (c), along the laser beams direction, with external magnetic field $B = 20$ T.
    }
    \label{fig:sim_results_LULI_Ne_5.8_7.2_7.8ns}
\end{figure}

The experiment was first modeled with the 3D MHD code FLASH (\citealt{FLASH}).
We model the formation and the propagation of pistons and shocks generated by the laser interaction with two Teflon targets having the same arrangement as shown in Fig.~\ref{fig:targets}. However, to reduce the computational cost, the separation between the targets is here limited to 6.5 mm, instead of 9 mm as in the experiment. As in the experiment, the targets are embedded inside an ambient hydrogen gas-jet within an external magnetic field. The laser intensity, the hydrogen gas-jet density, and the external magnetic field strength are the same as in the experiment.\\
Fig.~\ref{fig:sim_results_LULI_Ne_5.8_7.2_7.8ns} shows the electron density time evolution in the case with the external magnetic field (20 Tesla), i.e., before the shocks collision at t = 5.8 ns, at collision time t = 7.2 ns and after the collision at t = 7.8 ns, respectively. Due to the reduced distance between the targets used in the simulation, to scale it with the experiment, the collision time should be scaled by a factor 1.4, resulting in a scaled collision time of 10 ns, which is quite close to the experimentally observed one ($\sim 12$ ns). When the two shocks collide, the electron density increases only of $20 \%$. The pistons expand more slowly due to the increase of the magnetic pressure behind the shock. 
\\
These hydrodynamic simulations performed with FLASH are capable of describing the overall dynamic of the system, but they were not able to quantitatively reproduce the temperatures measured in the experiment (\citealt{yao2021laboratory}), likely due to the fact that the kinetic effects associated to our collisionless system cannot be taken into account.
That is why we have performed also PIC simulations to take them into account.

\subsection{PIC simulations}

The interaction between the two subcritical shocks has been modeled via the fully kinetic Particle-In-Cell code, SMILEI (\citealt{derouillat2018smilei}), for which we used the profiles of plasma density, temperature, and the magnetic field extracted from the experimental data as initial conditions (see Table~\ref{tab:params}).\\
We simulated such a system in a 1D3V geometry, as the scale of the shock front interaction with the ambient medium is much smaller across the shock (a few hundreds of microns) than along the shock (a few mm).
We point out that our PIC simulations are dedicated to capture only the kinetic effects of the shock colliding process. The laser-target ablation and piston formation are well-reproduced by the FLASH simulations and the shock formation and transition from supercritical to subcritical is detailed in our previous papers (\citealt{yao2021laboratory,yao2021detailed}).
\\
\begin{figure}[htp]
    \centering
    \resizebox{\hsize}{!}{\includegraphics[width=0.48\textwidth]{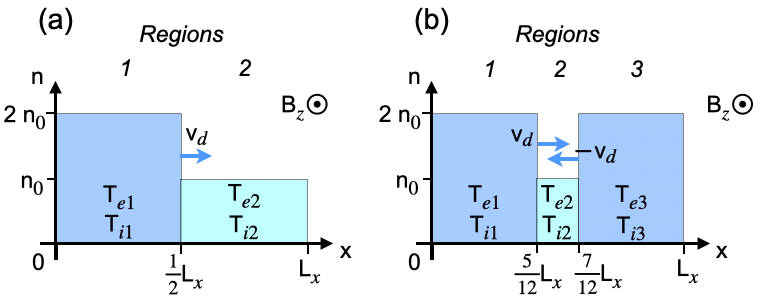}}
    \caption{1D PIC simulations initialization setups. (a) Single-shock case: a hydrogen plasma ($n_1 = 2n_0 = 2 \times 10^{18}$ cm$^{-3}$, $T_{e1} = 130$ eV and $T_{i1}  = 200$ eV) drifts through a background hydrogen plasma ($n_1 = n_0 = 1 \times 10^{18}$ cm$^{-3}$, $T_{e2} = 80$ eV, and $T_{i2} = 20$ eV) with a drifting velocity of $v_d = 350$ km/s (see Table~\ref{tab:PIC_sim_parameters0}). (b) Double-shock case: a background plasma ($n_2 = n_0 = 1 \times 10^{18}$ cm$^{-3}$, $T_{e2} = 80$ eV, and $T_{i2} = 20$ eV) is set at rest between two counter-streaming denser plasmas ($n_1 = n_3 = 2n_0= 2 \times 10^{18}$ cm$^{-3}$, $T_{e1} = T_{e3} = 130$ eV and $T_{i1} = T_{i3} = 200$ eV). The drifting velocity $v_d = 350$ km/s imposed to the protons in both configurations leads to a shock velocity of $v_s \approx 640$ km/s. 
    The simulation is initialized $\sim$ 11 ns after the lasers have started ablating the targets, with a distance of 1.8 mm in between the two shock fronts, while the box has a total length L$_x= 11$ mm.
    }
    \label{fig:1D_PIC_setup}
\end{figure}
In order to understand the effects of the collision of two shocks, both a single drifting shock configuration and a double counter-streaming shocks scenario have been simulated. 
The initial configurations are drawn in Fig. \ref{fig:1D_PIC_setup}.
The box has a length L$_x = 2048 \, d_e\approx 11$ mm and the spatial resolution is $d_x= 0.2d_e\approx 1.1$ \textmu m, where $d_e=c/\omega_{pe}\approx 5.3$ \textmu m is the electron inertial length, and $\omega_{pe}= \sqrt{n_{0} q^2_e/(m_e \epsilon_0)}\approx 5.6\times 10^{13}$ rad/s is the electron plasma angular frequency. Here, $c$ is the velocity of light, $n_{0} = 1.0 \times 10^{18}$ cm$^{-3}$ is the electron (and proton) number density of the ambient plasma, and $m_e$, $q_e$ and $\epsilon_0$ are the electron mass, the elementary charge, and the permittivity of free space, respectively. 
We point out that the x-axis we are talking about here in the case of PIC simulation does not correspond to the one relative to the experimental setup (used in Fig.~\ref{fig:double_shock_setup}, \ref{fig:targets}, \ref{fig:interf_2shocks}) and to the FLASH simulations (Fig.~\ref{fig:sim_results_LULI_Ne_5.8_7.2_7.8ns}). In our PIC simulation, the x-axis is the axis along which the two shocks propagate.
Each cell has 1024 particles plus 1 tracked particle for each species.
Moreover, an external uniform magnetic field $B_{z0} = 20$ T is set in the z-direction perpendicular to the plasma velocity ($\omega_{ce} / \omega_{pe} = 0.06$, where $\omega_{ce} = q_e B_{0z} / m_e$).
The simulation lasts for $1.5 \times 10^5 \omega_{pe}^{-1} \approx 2.5$ ns, with an initial time that corresponds to $\sim$ 11 ns in the experiment, i.e., $\sim$ 1 ns before the shocks collide. In short, the simulation covers time between 11 and 13.5 ns of the experiment. These times will be used below for better comparison with the experiment.
\\
We have initialized the system with different portions of plasma, all composed of protons and electrons with $m_p / m_e = 1836$.
For a single subcritical shock, we have set from 0 to L$_x$/2 (Region 1 in Fig.~\ref{fig:1D_PIC_setup}(a)) a hydrogen plasma drifting towards positive $x$ with a velocity $v_x = v_d = $ 350 km/s and density $n_1 = 2n_0 = 2\times 10^{18}$ cm$^{-3}$, while between L$_x$/2 and L$_x$ (Region 2) we put a background hydrogen plasma with density $n_2 = n_0$ at rest.
In the case of double shock, we set two counter-streaming hydrogen plasmas with densities $n = 2n_0$ moving in the $x$-direction at velocities $v_d$ and $-v_d$ between 0 and 5 L$_x$/12 (Region 1 in Fig.~\ref{fig:1D_PIC_setup}) and between 7 L$_x$/12 and L$_x$ (Region 3), respectively.
Moreover, a background hydrogen plasma with density $n_0$ was set at rest in between (Region 2).
As for the temperature, we have used the results of the TS diagnostic in the experiment for both simulations, i.e., $T_{e1} = T_{e3} = 130$ eV and $T_{i1} = T_{i3} = 200$ eV for the drifting plasmas, and $T_{e2} = 80$ eV and $T_{i2} = 20$ eV for the background plasma. 
The list of parameters used to initialize our simulations is summarized in Table~\ref{tab:PIC_sim_parameters0}.
\begin{table}[htp]
	\begin{center}
	\caption{List of parameters we have used to initialize our PIC simulations. Regions 1 and 3 correspond to the drifting hydrogen plasmas, while Region 2 is relative to the background hydrogen plasma at rest. 
    $\lambda_{mfp}^{i-i \,(d-a)}$ is the mean free path relative to the collisions between drifting and ambient ions, while $M_{ms}$ is the magnetosonic Mach number of the shock wave moving at speed $v_s = 640$ km/s in the ambient plasma characterized by a magnetosonic speed $c_{ms}= 448$ km/s. We point out that the values relative to the Regions 1 and 3 refer only to the initial situation, before the shocks completely form, hence they are not to be confused with the parameters of the downstream region at a certain time.
    }
    \label{tab:PIC_sim_parameters0}
    \begin{tabular}{c|c|c}
    \hline
    \multirow{2}{*}{\shortstack{\textbf{PIC initialization} \tabularnewline \textbf{parameters}}} 	&  \multicolumn{2}{c}{\textbf{Regions}} \tabularnewline
   	&  \textbf{1 and 3}	& \textbf{2} \tabularnewline
    \hline
     $v_d$ [km/s]	 	& 	
     350	 & 0 \tabularnewline
     $v_s$ [km/s] 	&  
     640  & 0 \tabularnewline
     $B_{z0}$ [T]	&	20  &  20  	\tabularnewline
     $T_e$ [eV]  & 130 &  
     80 \tabularnewline
     $T_i$ [eV]  & 200  &  
     20 \tabularnewline
     $n_i$ [$10^{18}$ cm$^{-3}$]   & 2  &  1 	\tabularnewline
     $v_A$	[km/s]  &	308  & 436				\tabularnewline
     $c_s$ [km/s]   &   144    &   102 \tabularnewline
     $c_{ms}$ [km/s] &  340    &   448 \tabularnewline
     $r_{Li}$	[mm]	&	0.26  &   0.39	 \tabularnewline
    $\lambda_{mfp}^{i-i \, (d-a)}$ [mm] &	\multicolumn{2}{c}{33}	 \tabularnewline
     $M_{ms}$ 	&	\multicolumn{2}{c}{     1.43}     \tabularnewline
     \hline
    \end{tabular}
    \end{center}
\end{table}
\begin{figure}[htp]
    \centering
    \includegraphics[width=0.48\textwidth]{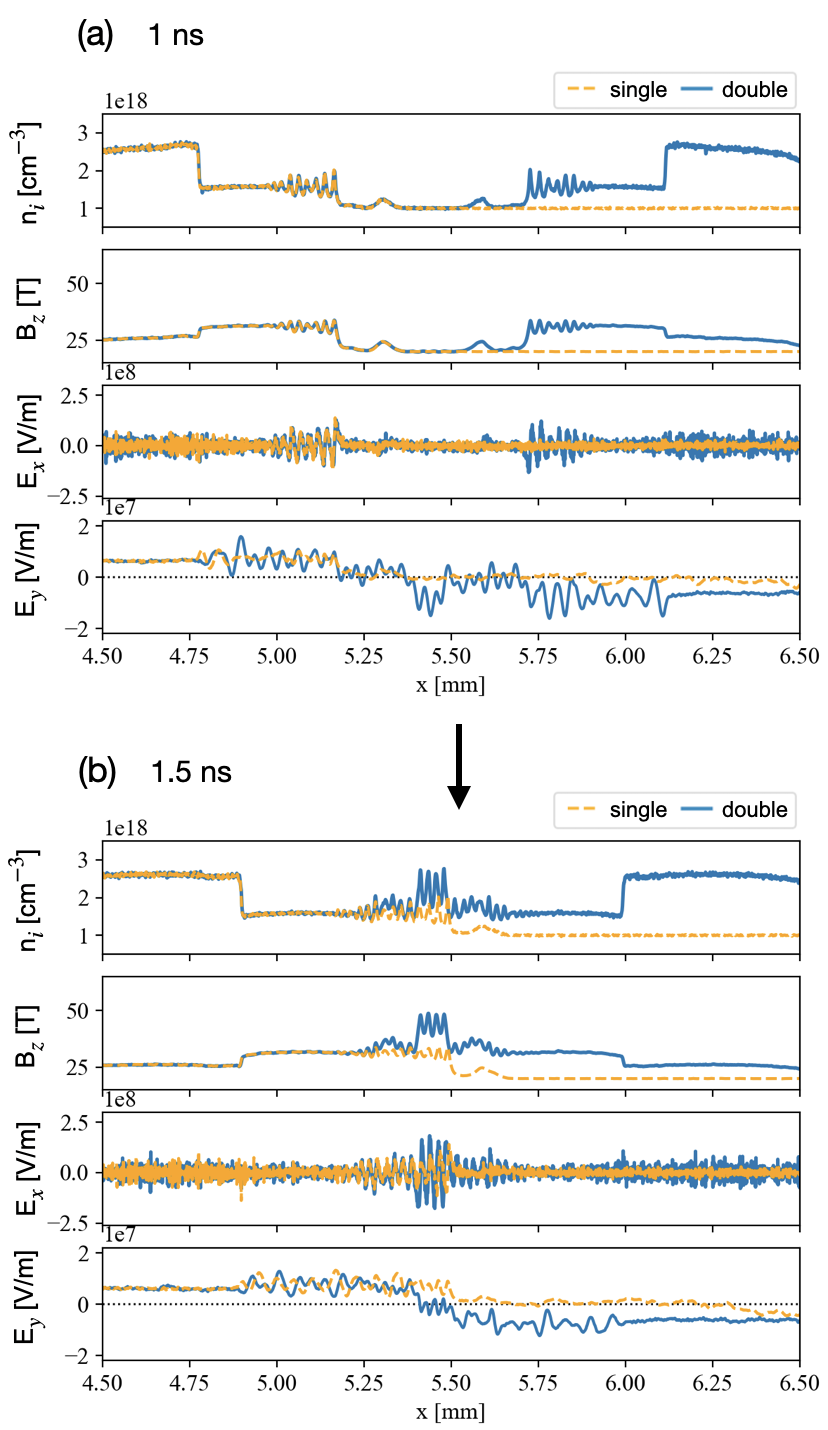} 
    \caption{Profiles of ion density, magnetic field $B_z$, and electric fields $E_x$ and $E_y$ at times 1 ns (a) and 1.5 ns (b) after the beginning of the simulation. At 1 ns ($\approx 12$ ns for the experiment), the two fast shocks propagate without perturbing each other, as their profiles correspond to the one of the single shock case. 
    At 1.5 ns ($\approx 12.5$ ns for the experiment), the interaction between the two shocks makes the ion density and the magnetic field $B_z$ increase and pile up in the middle; 
    the electric field $E_x$ presents several spikes also due to the interaction between the downstream fast magnetosonic waves;
    the inductive $E_y$ does not drastically change structure or magnitude, but it presents opposite signs in the downstream regions depending on the propagation direction of the shock. We point out that the profiles relative to the single shock have been shifted of $\Delta x = -$ L$_x/12$ in order to simplify the comparison.}
    \label{fig:1vs1,5ns_profiles}
\end{figure}

Fig.~\ref{fig:1vs1,5ns_profiles} shows the spatial profiles of ion density, magnetic field $B_z$, and electric fields $E_x$ and $E_y$, for both single shock and double shock simulations. We point out that the profiles relative to the single shock have been shifted by $\Delta x = -$ L$_x/12$ in order to simplify the comparison.
In Fig.~\ref{fig:1vs1,5ns_profiles} (a) at 1 ns ($\approx 12$ ns for the experiment), we clearly see that the shock formation and propagation happens in the same way for both the single shock case and the double shock case, as the profiles relative to the single shock overlap the ones of the right-drifting shock in the double shock case.
In other words, the presence  of the left-drifting shock has no effect on the evolution of the right-drifting one yet and vice versa.
In the downstream regions, we note the spontaneous formation of fast magnetosonic waves propagating away from the shock fronts (\citealt{Moreno2019}).
We point out that in the double-shock case the electric field components $E_x$ and $E_y$ of the right-drifting shock and the left-drifting one have opposite direction: the shock coming from the left is characterized by oscillations starting with a positive peak of $E_x$ at the shock front and by a downstream region with $E_y>0$, while the shock coming from the right has oscillations starting with a peak $E_x<0$ at the shock front and a field $E_y<0$ in the downstream region. 
\\
In Fig.~\ref{fig:1vs1,5ns_profiles} (b) at 1.5 ns ($\approx$ 12.5 ns for the experiment), the interaction of the two shocks has begun: the ion density and magnetic field start overlapping, and, at longer times not shown here, they keep piling up, reaching values of $n_i \approx 5 \times 10^{18} $cm$^{-3} = 5 \, n_0$ and $B_z \approx 50$ T $= 2.5 \, B_{z0}$ at $\sim 2.2$ ns ($\approx 13$.2 ns in the experiment). 
The $x$-component of the electric field $E_x$ fluctuates around 0 and has peaks of $\sim$ 100 MV/m associated with the shocks. After the interpenetration of the two shocks, $E_x$ reaches values of $\sim$ 250 MV/m.
The only contributions to the $y$-component $E_y$ is the inductive electric field $E_y=-(\mathbf{v}\times \mathbf{B})_y = v_xB_z$ (\citealt{ilie2017}) and it, too, has a fluctuating profile, centered on 0 in the upstream regions and on $\sim \pm$ 8 MV/m in the downstream ones. Between these two zones, $E_y$ passes gradually from 8 MV/m (or $-$8 MV/m) to 0, as the plasma velocity distribution decreases (increases) and  the magnetic field increases (decreases), even after the two shocks have met.

\begin{figure}[htp]
    \centering
    \resizebox{\hsize}{!}{\includegraphics[width=0.48\textwidth]{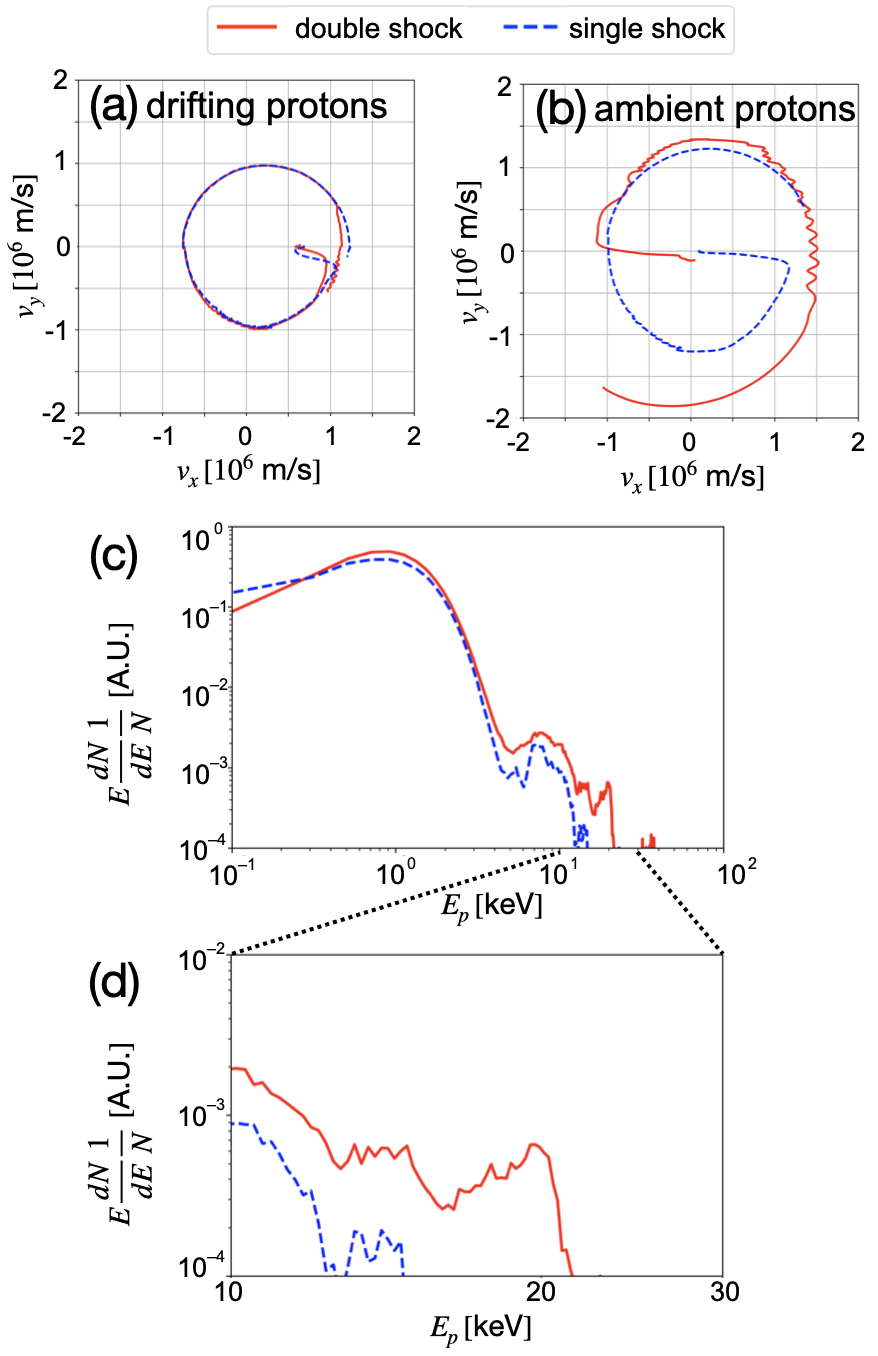}}
    \caption{Comparison of the trajectories in the $v_x-v_y$ space of two typical energetic tracked drifting (a) and ambient (b) protons, for the double shock case in red and for the single shock case in blue.
    (We note that their initial velocities might be different from the species bulk speed ($v_x = 350$ km/s), since the protons are initiated with an initial temperature.)
    (c) Final energy spectra of the ions for the two configurations and (d) zoom on the range from 10 to 30 keV.
    }
    \label{fig:vx_vy_spectra}
\end{figure}

\begin{figure}[htp]
    \centering
    \includegraphics[width=0.48\textwidth]{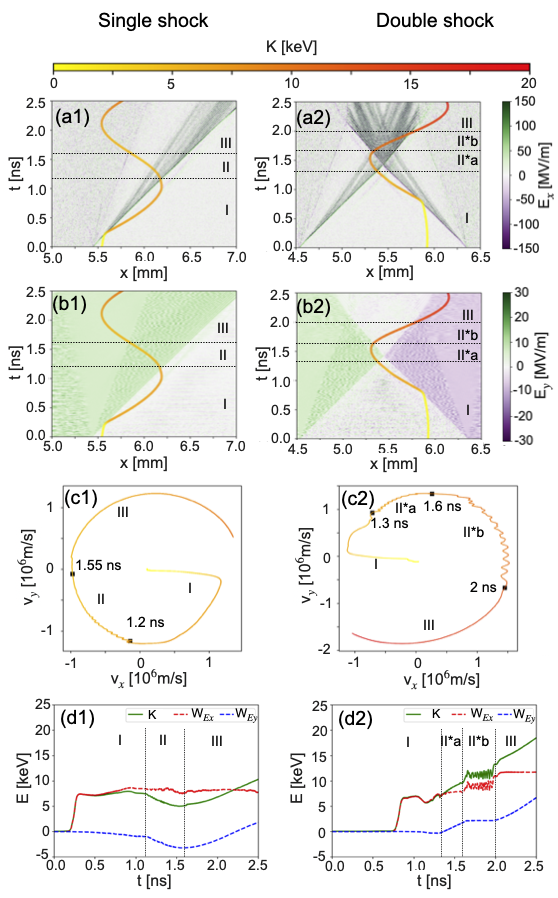}
    \caption{Comparison of the trajectories of two typical energetic protons of the ambient plasma in the single- and double-shock cases in the left column (1) and right column (2), respectively.
    In their evolution between 0 and 2.5 ns we can distinguish different phases (I, II, II*a, II*b, III), that are detailed in the main text. In (a1-a2) and (b1-b2), the maps of the $E_x$ and of the $E_y$ fields in the x-t space, together with the trajectories of the chosen protons, are shown, respectively. Note the different scales for the electric fields $E_x$ and $E_y$ color maps. In (c1-c2), the proton trajectories in the vx-vy space and the temporal points delimiting the different phases, are plotted. The color of the proton trajectories in (a1-a2), (b1-b2), and (c1-c2) follows the evolution of their kinetic energy $K$ on a color scale from 0 to 20 keV. In (d1-d2), are shown the temporal evolution of the kinetic energies of the two tracked protons (full green line), the work done by $E_x$ (dashed red line) and by $E_y$ (dashed blue line).}
    \label{fig:SS_DS_500}
\end{figure}

By tracking a set of representative protons, we have been able to understand the energization mechanism undergone by the most energetic ones, i.e., the ones reaching a kinetic energy of $E_k > 10$ keV.
In Fig. \ref{fig:vx_vy_spectra} (a) and (b) we show the motion in the $v_x$-$v_y$ space of protons from the drifting and the ambient plasma, respectively. For clarity, we have plotted for each plasma only one particle for each case, but we have checked that these trajectories are well representative of all other tracked particles coming from the same populations.
We observe that the protons of the drifting plasma, having initially a bulk velocity $v_x = 350$ km/s, rotate in the $v_x$-$v_y$ space, without showing any special difference between the single and double shock cases.\\
The situation is definitely different for the protons of the ambient plasma: after starting at rest, the protons are accelerated by the shocks up to kinetic energies that, in the double shock case,
are around  1.5 times higher than the single shock case.
This difference is well presented by the proton spectra at the final simulation time shown in Fig.~\ref{fig:vx_vy_spectra} (c) and, zoomed, in Fig.~\ref{fig:vx_vy_spectra} (d), where to higher energies are associated higher distribution values in the double-shock case.\\
By comparing the cases with one and two subcritical shocks, we could understand the reason of the higher energization in the double shock case by analyzing the dynamics of such energetic tracked particles.
Let us start considering the protons of the ambient plasma whose dynamics is reported in Fig. \ref{fig:vx_vy_spectra} (b). 
For the proton from the single shock case, we have analyzed its motion in the in $x$-$t$ space over a map of $E_x$ and a map of $E_y$ (Fig. \ref{fig:SS_DS_500} (a1) and (b1), respectively), and in the $v_x$-$v_y$ space (Fig. \ref{fig:SS_DS_500} (c1)), where the color of the proton trajectory follows a scale based on its kinetic energy $K$. Moreover, we have compared the temporal evolution of the proton kinetic energy and the work done by $E_x$ and $E_y$ on it as shown in Fig. \ref{fig:SS_DS_500} (d1). On these four graphs (a1-d1) we have distinguished the presence of three phases corresponding to different regimes experienced by the proton. 
\\
In phase I, the proton is accelerated by the electrostatic field $E_x$ associated with the shock front, it gains a velocity $v_x>0$ and starts rotating in the upstream plasma.
While gyrating clockwise due to the applied magnetic field $B_z$, its velocity in the $x$-direction $v_x$ decreases and the proton meets again the shock front. After crossing it, in phase II, the proton is in the downstream region characterized by $E_y>0$. Since it has a $v_y<0$, the positive $E_y$ does a negative work on the proton and lowers its kinetic energy, as can be seen in Fig.~\ref{fig:SS_DS_500} (d1).
In phase III, the particle has again a velocity $v_y>0$, hence the positive $E_y>0$ does a positive work on the proton and keeps energizing it.
\\
We have conducted a parallel analysis on the proton from the ambient plasma in the double shock case and plotted the results on the right column of Fig. \ref{fig:SS_DS_500}: (a2) and (b2) show its motion in the x-t space over a map of $E_x$ and $E_y$, respectively; (c2) shows its trajectory in the $v_x$-$v_y$ space; (d2) shows the temporal evolution of the particle kinetic energy and the work made on it by $E_x$ and $E_y$. The color of the proton trajectory in Fig. \ref{fig:SS_DS_500} (a2)-(c2) follows the same kinetic energy scale used for the single shock case.
We can now distinguish four different regimes for the tracked proton.
In phase I, similarly to what happens in the single shock case, the proton is accelerated by the right shock front and rotates in the upstream region.
Phase II*a and II*b are due to the presence of the second shock (coming from the right side) and hence are different to phase II in the single shock scenario.
Specifically, in phase II*a, the proton interacts with the shock front coming from the left, whose $E_x$ oscillates but does a net positive work on the proton, increasing its energy.
In phase II*b, the proton encounters the shock coming from the right and finds itself in the region downstream of both shocks, which is characterized by a low $E_y$ (due to the colliding of the two shocks) and by an oscillating $E_x$ (due to the interaction of the two magnetosonic waves), that do not change its energy in a net way.
Phase III in the double shock case is similar to the one for the single shock: the proton is in the downstream zone of the shock coming from the right and has a velocity $v_y<0$, hence the negative $E_y<0$ makes a positive work and keeps energizing it.
\\
Other energetic protons show a similar behavior: in general, the higher energization in the double shock case is mostly due to the fact that, while gyrating because of the magnetic field, the proton can find areas with $E_y$ directed accordingly to its velocity $v_y$ or zones downstream (of both shocks) with low $E_y$ that does not decelerate it either. 
These two contributions have the effect of accelerating some of the ambient ions to higher energies in the case of two counter-propagating shocks than in the presence of only one.

Among the tracked background ambient protons, we can compare how many reach, for instance, at least 10 keV in the cases of single and double shock.
In the presence of a single shock only 1 proton out of 5120 ($\approx 0.02 \%$) reaches 10 keV or higher energies, while with the second shock we have 20 out of 1706 ($\approx 1.17 \%$), hence a much higher percentage manages to be accelerated to higher energies. We note that the inequality of number of tracked protons in the two cases comes from the difference of ambient plasma size.

We point out that we are not able to compare the experimental proton spectrum with the simulated one: during the first 3-4 ns of evolution, the shocks are supercritical and the individual interaction of both of them with the background plasma already leads to some high particle energization, as shown in our previous work (\citealt{yao2021laboratory,yao2021detailed}).
As the shocks propagate, their velocity drops, leading to an interaction between two subcritical shocks. This interpenetration is indeed the only part that we are simulating here, hence, the resulting spectrum cannot take into account the protons previously accelerated by the supercritical shocks.
The protons accelerated by supercritical shocks reach much higher energies, hence in the final spectrum obtained experimentally they ``cover'' the portion of protons accelerated to lower energies by the interaction between the two subcritical shocks.\\

We have considered, as detailed in the Appendix, what is the limit of validity of the 1D framework used here.

\section{Astrophysical relevance}

Although the vast majority of collisionless shocks in astrophysics are supercritical, the results of our experiment can be relevant for some phenomena observed in both interplanetary and astrophysical plasmas and involving subcritical shocks. In fact, observations of subcritical shocks are sparse and most of them are restricted to the interplanetary space, where colliding shocks can also be observed (e.g., \citealt{1966SSRv....5..439C}). Examples are the interplanetary forward shocks convected with the solar wind that are expected to propagate with the wind mostly outward into the outer heliosphere, so that they can have a sufficiently low Mach number to be subcritical. Furthermore, subcritical shocks are expected in cometary environments, where the collision between the solar wind and the atmosphere of the comet can reduce the upstream flow velocity, resulting in low Mach number cometary bow shocks. CMEs can also produce subcritical shocks. The observations show that CMEs resulting from x-ray flares of the solar corona associated with type-II radio bursts (the so-called radio-loud CMEs) are, in general, very fast and extended and, initially, they produce supercritical shocks that become subcritical at later times (e.g., \citealt{bemporad2011identification, bemporad2013super}); CMEs resulting from flares not producing a type-II burst (radio-quiet CMEs) produce subcritical shocks at all times (e.g., \citealt{bemporad2011identification, bemporad2013super}). All these subcritical shocks are present in the interplanetary space and can interact with each other or with planetary bow shocks, thus contributing to the acceleration of particles (electrons, protons, ions) up to near-relativistic energies. These shock-shock interactions are reproduced by our experiment which shows that ambient ions can be energized around 1.5 times more than in single shocks and that the particles are accelerated in different ways, depending on the areas of the shock-shock interaction region, where the particles are located. The implication is that the population of high-energy particles in the interplanetary space may depend on the rate of occurrence of shock-shock interactions.

Thanks to the Voyager 1 and 2 missions, since 2012 it was possible to probe the density of the very local interstellar medium with accurate in situ measurements. This has allowed to unveil the presence of several shock waves in the interstellar plasma, most likely interplanetary shocks originated from energetic solar events (e.g., CMEs) that traveled outward through the supersonic solar wind and, after colliding with the heliospheric termination shock, crossed through the heliopause into the interstellar medium (\citealt{2013ApJ...778L...3B, 2013Sci...341.1489G, 2014ApJ...788L..28L}).
\cite{2013ApJ...778L...3B} reported the first in situ measurement of a shock in interstellar plasma, whose characteristics are summarized and compared to the ones relative to our shocks in Table~\ref{tab:comparison_shock_Burlaga}. Then evidence of multiple shocks was reported in 2015 data collected with Voyager 1 (\citealt{2021NatAs...5..761O}). According to the Voyager measurements these shocks are, in general, weak low beta and subcritical shocks (\citealt{1984JGR....89.2151M, 2013ApJ...778L...3B, 2018ApJ...854L..15M}). Thus we can argue that the results of our experiment may be applicable to the interactions between these subcritical shocks which populate the very local interstellar medium. Interestingly, these shocks of solar origin are characterized by a precursor consisting of various disturbances in the intensity and anisotropy of galactic cosmic rays (\citealt{2015ApJ...809..121G}). Voyager missions have revealed that these disturbances are typically preceded by bursts of high-energy ($\approx 5-100$~MeV) electrons, most likely due to the reflection and acceleration of cosmic-ray electrons by magnetic field jumps at the shock and/or due to interactions with upstream plasma waves/shocks (\citealt{2021AJ....161...11G}). Our experiment shows that the interaction between these subcritical shocks has a direct effect on the way particles are accelerated by the shocks and on the maximum energization of particles.

Possible shock-shock interactions as those discussed above can also be present (and play a significant role in the acceleration of particles) in the environments around exoplanets. In addition to cases analogous to those we observe in our solar system, the cases of hot-Jupiters (gas giant exoplanets that should be similar to Jupiter but are in close proximity to their stars) are of particular interest, given the strong interaction with their host stars via the stellar wind, the magnetic field and the irradiation. Depending on the parameters of the star-planet system (distance between the two objects, masses of the star and the planet, wind velocity, stellar irradiation, etc.) complex flow structures can form from the colliding planetary and stellar winds, as bow shocks, cometary-type tails, and inspiraling accretion streams (e.g., \citealt{Matsakos2015}). In particular, the speed of the planetary wind is, in general, only marginally supersonic, so that the Mach number is low (e.g., \citealt{2013MNRAS.428.2565T}). Also the cometary-type tails around the exoplanets are advected by the stellar winds (so that possible shocks can have low Mach numbers to be subcritical) and can be highly perturbed (producing a highly variable complex pattern of shocks), depending on the parameters of the star-planet system. Under these conditions, interactions between subcritical shocks may develop and can be analogous to those produced in our experiment.

\begin{table}[htp]
    \begin{center}
       \caption{Dimensionless quantities relative to our laser-driven shock and the interstellar medium shock characterized by Voyager in \citealt{2013ApJ...778L...3B}, as a weak subcritical resistive laminar shock. $DS$ and $US$ refer to the downstream and upstream zones, respectively.}
    \label{tab:comparison_shock_Burlaga}
    \begin{tabular}{c|cc}
    \hline
    \textbf{Parameters}    & \textbf{Our shocks}   & \textbf{Shocks as in (1)} \\
    \hline
    $B_{DS}/B_{US}$         & 1.5   &  1.4  \\
    $\beta_{ther, US}$      & 0.1   &  0.23 \\
    $(v_A/c_s)_{US}^2$      &  4.6  &  5    \\
    $\theta_{Bn}$           &  90   &  85   \\
    $M_{ms}$            &  $1.12^{\textrm{(exp)}}-1.43^{(\textrm{sim})}$   & $\approx1.9$   \\
    \hline
    \end{tabular}
    References: (1) \cite{2013ApJ...778L...3B}.
    \end{center}
\end{table}

\section{Conclusions}
In our experimental campaign we have investigated the interpenetration of two laser-driven collisionless subcritical shocks, relevant for astrophysical phenomena such as the interplanetary medium and the local interstellar medium.
The data obtained from their characterization has been used to feed MHD and PIC simulations, respectively run with the FLASH and the SMILEI codes. While the MHD simulations have been used to describe the collisional part of the problem, i.e., the piston, and the overall evolution of the system; PIC simulations have provided insights into the microphysics at play in such a scenario. We compared the cases of single and double shocks and observed an acceleration of the background ions which was up to 1.5 times higher in the case of double shocks. Such acceleration is initially due to the electrostatic field $E_x$ associated with the shock front (injection), followed by a contribution mostly due to the ``surfing'' effect on the inductive electric field ($E_y \sim v_x B_z$). The presence of a second shock benefits this second mechanism as it allows the existence of zones with $E_y$ directed as the proton $v_y$, and thus enhance their acceleration.
Unfortunately we could not measure such a spectrum during our experimental campaign, as the previously formed supercritical shocks produced higher energy protons whose spectrum covered the one of the particles accelerated by the further stage of double subcritical shock interaction. \\
In spite of the fact that most astrophysical collisionless shocks are supercritical, these results can shed light on the less investigated subcritical shocks which are still relevant in various space phenomena. In particular, we have shown that the interaction of two subcritical shocks could lead to a higher energization of the background protons. This is a relevant information when determining the distribution of high-energy particles that populate the interplanetary space and the very local interstellar medium surrounding the heliopause where colliding subcritical shocks are present.
Moreover, high-power laser-plasma experiments have demonstrated to be an essential tool allowing us to recreate in the laboratory scaled astrophysical phenomena, whose characterization is crucial to initialize the relative numerical simulations.

\begin{acknowledgements}
The authors would like to thank the teams of the LULI2000 (France) laser facility for their expert support, as well as the Dresden High Magnetic Field Laboratory at Helmholtz-Zentrum-Dresden-Rossendorf for the development of the pulsed power generator. We thank the SMILEI development team for technical support; we thank P. Loiseau (CEA-France) for the Thomson scattering analysis code. This work was supported by funding from the European Research Council (ERC) under the European Unions Horizon 2020 research and innovation program (Grant Agreement No. 787539). 
S.O. and M.M. acknowledge financial contribution from the PRIN INAF 2019 grant ``From massive stars to supernovae and supernova remnants: driving mass, energy and cosmic rays in our Galaxy'’ and the INAF mainstream program ``Understanding particle acceleration in galactic sources in the CTA era''.
The computational resources of this work were supported by GENCI. The FLASH software used in this work was developed in part by the DOE NNSA ASC - and DOE Office of Science ASCR-supported FLASH Center for Computational Science at the University of Chicago. Part of the experimental system is covered by a patent (1000183285, 2013, INPI-France). The work of JIHT RAS team was supported by The Ministry of Science and Higher Education of the Russian Federation (Agreement with Joint Institute for High Temperatures RAS No 075 15 2020 785).
\end{acknowledgements}

\bibliographystyle{aa}
\bibliography{main}

\section*{Appendix A: limits of validity}

In our 1D PIC simulations we have not taken the multi-dimensional effects into consideration, e.g., the shock front non-stationarity (\citealt{burgess2007shock}), which might affect the proton dynamics due to the rippling along the shock front (\citealt{yang2012impact}). 
However, we have verified in our former work that the proton acceleration mechanism in the single shock case is not affected by the non-stationarity in the early few ns (\citealt{yao2021detailed}). 

Moreover, the 1D geometry approximation has the limit of not considering the hemispherical profile of the shock. This can become a problem if the particles have gone too far in the y-direction and exit the shock.
To quantify the maximum length that the protons can travel, we consider the radius and the thickness of the shock right before the collision: from the experimental characterization obtained via interferometry, we estimate a shock front radius $R \approx 2.5$ mm and a shock thickness $\delta R \approx 0.2$ mm.

\begin{figure}[htp]
    \centering
    \resizebox{\hsize}{!}{\includegraphics[width=0.4\textwidth]{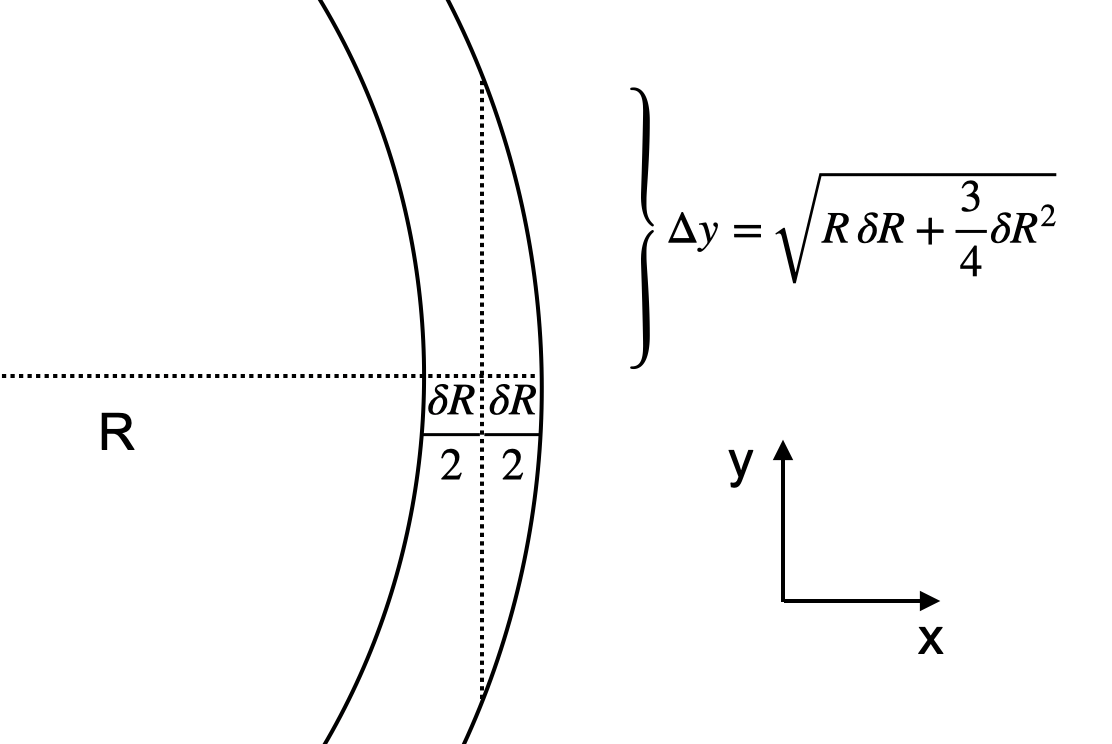}}
    \caption{Estimation of the maximum distance that a particle can travel in the y-direction before escaping the shock, considering the hemispherical geometry of the shock front. If we consider a radius $R\approx 2.5$ mm and a shock thickness $\delta R \approx 0.2$ mm, we have $\Delta y \approx 0.73$ mm.}
    \label{fig:1D_limits}
\end{figure}

As shown in Fig.~\ref{fig:1D_limits}, we approximate the maximum length $\Delta y$ for which the particle can still be considered inside the shock as the distance that a particle moving only along the y-axis would travel between the middle of the shock and the external edge of it. This length would result in $\Delta y = \sqrt{R \, \delta R + \frac{3}{4} \delta R^2} \approx 0.73$ mm.\\
This leads to imposing such a limit to the background protons that have been accelerated by the double shock. Among the 20 protons with energies of 10 keV and higher, only 8 ($=40 \%$) manage to remain confined in the shock all the time. This means that out of the total 1706 tracked ambient protons, only 8 ($\approx 0.47 \%$) reach at least 10 keV while remaining confined in the shock.
Hence, even considering this limitation, the percentage of protons that are energized to 10 keV and more is considerably higher in the presence of a double shock structure than with a single one (1 out of 5120, i.e., $\approx 0.02 \%$).\\

\end{document}